\documentclass[aps,prl,reprint,superscriptaddress,nofootinbib]{revtex4-2}

\usepackage{amsmath,amssymb,bm}
\usepackage{graphicx}
\usepackage{xcolor}
\usepackage{hyperref}
\graphicspath{{figures/}} 

\begin{document}

\title{Hidden chirality and half-vortex formation in polariton condensates}

\author{A. N. Osipov}
\affiliation{School of Physics and Engineering, ITMO University, Saint Petersburg 197101, Russia}
\author{I. Y. Chestnov}
\affiliation{School of Physics and Engineering, ITMO University, Saint Petersburg 197101, Russia}
\author{P. G. Lagoudakis}
\affiliation{Hybrid Photonics Laboratory, Skolkovo Institute of Science and Technology, Bolshoy Boulevard 30, Building 1, Moscow 121205, Russia}
\author{A. V. Yulin}
\affiliation{School of Physics and Engineering, ITMO University, Saint Petersburg 197101, Russia}

\date{April 29, 2026}

\begin{abstract}
We show that a radially symmetric, nonresonantly pumped spinor exciton-polariton condensate can acquire chirality without a rotating drive, chiral geometry, or pump orbital angular momentum. Spin relaxation in a two-reservoir system shifts the reservoir-induced blueshift relative to the gain profile, creating an effective non-Hermitian chiral potential. A reduced angular-mode theory reveals tunable exceptional points and non-reciprocal coupling between counter-rotating modes. Full driven-dissipative Gross-Pitaevskii simulations show that this hidden chirality enables all-optical formation of spin-selective half-vortices.
\end{abstract}

\maketitle

Topology and chirality provide organizing principles for a wide range of collective phenomena in condensed matter. Quantized vortices are the defining excitations of superfluids and superconductors, from Abrikosov flux lattices in type-II superconductors~\cite{Abrikosov1957} to rotating atomic Bose-Einstein condensates~\cite{AboShaeer2001}, while fractional and spin-textured defects such as half-quantum vortices and skyrmions encode additional internal degrees of freedom of multicomponent order parameters~\cite{Sondhi1993,Nagaosa2013,Autti2016}. In exciton-polariton condensates, the driven-dissipative nature of the fluid, the spinor structure inherited from circularly polarized cavity modes, and the strong coupling to incoherent reservoirs provide a distinct setting in which topology, gain, loss, and interactions coexist on microscopic time scales~\cite{Kasprzak2006,Sanvitto2016}. Half-vortices, predicted as elementary topological defects of spinor polariton condensates~\cite{Rubo2007} and observed experimentally in semiconductor microcavities~\cite{Lagoudakis2009}, are particularly attractive because their orbital angular momentum is tied to polarization, offering a route to spin-selective control of circulation.

Realizing this potential requires a mechanism that both creates vortices, and selects the sign and polarization content of the condensate vorticity. In most existing approaches this selection is imposed externally, for example through chiral polaritonic lenses~\cite{Dall2014}, rotating optical potentials~\cite{Gnusov2023,Redondo2023}, or asymmetric geometries~\cite{Sedov2021}. Here we demonstrate a different route: chirality can be generated dynamically in a stationary annular trap whose total pump intensity is radially symmetric and carries no orbital angular momentum. The required handedness is hidden in the reservoir response. A spin-dependent azimuthal modulation of the nonresonant pump, combined with spin relaxation in a two-component exciton reservoir, shifts the conservative blueshift landscape relative to the gain landscape. The condensate therefore experiences an effective non-Hermitian potential that is chiral even though no rotating external drive is applied.

This hidden chirality can become extreme near exceptional points (EPs), where counter-rotating angular modes coalesce and their coupling becomes effectively one-way~\cite{Dembowski2003,Peng2016,Ashida2020}. By tuning the relative phase and amplitude of the spin-dependent pump modulation, the EPs can be addressed selectively in one circular polarization. We show that this produces half-vortex states, in which one circular polarization has unit phase winding while the other remains in a zero-winding state~\cite{Rubo2007,Lagoudakis2009,pukrop2020circular}.
Our approach provides a deterministic pathway for half-vortex formation in non‑equilibrium polaritonic condensates, with full control over both the vorticity direction and the vorticity polarization, unlike previous works on half‑vortices in quasi‑conservative ~\cite{Lagoudakis2009, Rubo2007} or in non‑equilibrium high‑excitation‑power regimes~\cite{pukrop2020circular}, where formatting is spontaneous and probabilistic.

The condensate formation is described by a driven-dissipative Gross-Pitaevskii equation for the spinor order parameter $\Psi=(\psi_+,\psi_-)^\mathrm{T}$ coupled to inactive, $X^{(I)}_\pm$, and active, $X^{(A)}_\pm$, reservoirs of incoherent excitons~\cite{Gnusov2020}:


\begin{subequations}\label{eq.2DGPE}
\begin{eqnarray}
    \label{eq:gpe}
    &&i\hbar\partial_t \psi_{\pm} = \Big[-\frac{\hbar^2\bm{\hat{\nabla}}^2}{2 m_p}  + g(X_\pm^A + X_\pm^I) +\alpha|\psi_{\pm}|^2 \\
    &&~~~~~~~~~~~~~+\frac{i\hbar}{2}(RX_{\pm}^A - \Gamma)\Big]\psi_\pm, \notag \\
    \label{eq:active}
    &&\partial_t X_{\pm}^A = WX^I_{\pm} - (R|\psi_\pm|^2 + \Gamma_A)X_\pm^A  \\ \notag
    &&~~~~~~~~~~~~~ -\Gamma_s (X_\pm^A - X_\mp^A), \\
    &&\partial_t X_{\pm}^I = P_{\pm} - (W + \Gamma_I)X_\pm^I - \Gamma_s (X_\pm^I - X_\mp^I).
    \label{eq:inactive}
\end{eqnarray}
\end{subequations}

Here $m_p$ is the polariton effective mass; $\alpha$ and $g$ are the condensate self-interaction and reservoir-induced blueshift constants; $R$ is the stimulated scattering rate from the active reservoir; $W$ is the conversion rate from inactive to active excitons; $\Gamma$, $\Gamma_A$, and $\Gamma_I$ are the decay rates of the condensate, active reservoir, and inactive reservoir, respectively; and $\Gamma_s$ is the spin-relaxation rate. For clarity, we use the same spin-relaxation rate in the active and inactive reservoirs; the mechanism persists for unequal rates provided that spin relaxation induces a finite angular displacement between the reservoir components.

\begin{figure*}[t]
\includegraphics[width=\textwidth]{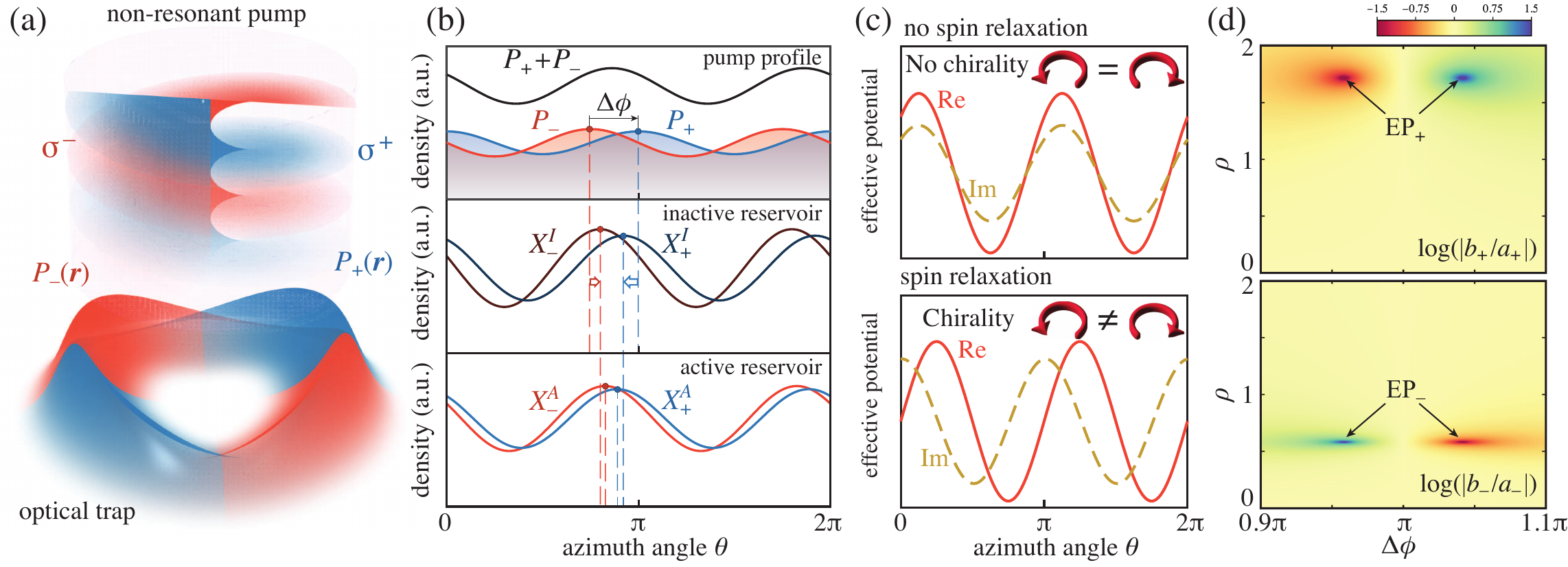}
\centering
\caption{Reservoir-mediated hidden chirality in an annular polariton trap. (a) Ring-shaped nonresonant pump with spin-dependent azimuthal intensity modulation in the circular components $\sigma^+$ and $\sigma^-$. The total pump intensity remains radially symmetric and carries no imposed orbital angular momentum. (b) Spin relaxation shifts and smooths the inactive and active reservoir profiles relative to the corresponding pump components. (c) Because inactive excitons contribute only to the conservative blueshift whereas active excitons contribute both blueshift and gain, the real and imaginary parts of the effective potential acquire a relative angular phase shift. This converts a non-rotating spin modulation into a chiral non-Hermitian potential. (d) Calculated non-reciprocity ratio between the effective couplings of counter-rotating modes as a function of the pump-modulation amplitude ratio $\rho=|p_2^-/p_2^+|$ and relative phase $\Delta\phi$. Singular points indicate exceptional points (EPs) where one of the directional coupling amplitudes vanishes.}
\label{fig:hidden_chirality}
\end{figure*}

We consider an annular pump with a weak second angular harmonic, schematically depicted in Fig.~\ref{fig:hidden_chirality}(a),
\begin{equation}
P_\pm(r,\theta)= f(r)\left[P_0+p_{2}^{\pm}e^{2i\theta}+\left(p_{2}^{\pm}\right)^*e^{-2i\theta}\right],
\label{eq:pump}
\end{equation}
where
\begin{equation}
f(r)=\frac{d^4}{\left(r^2-R_0^2\right)^2+d^4}.
\end{equation}
with $R_0$ being a trap radius, and $d$ being a trap wight.

The coefficients $p_0$ and $p_2^\pm=|p_2^\pm|e^{i\phi_\pm}$ are chosen so that $P_\pm(r,\theta)\geq0$. We denote the relative phase and amplitude ratio of the two spin components by
\begin{equation}
\Delta\phi=\phi_- -\phi_+,\qquad \rho=\left|\frac{p_2^-}{p_2^+}\right|,
\label{eq:rho}
\end{equation}
using $\rho$ for the pump-modulation ratio. 
The pump contains no azimuthal phase winding; its angular dependence is an intensity modulation proportional to $\cos(2\theta+\phi_\pm)$, shown in Fig.1(b).

The physical origin of the effect is the two-reservoir structure in Eqs.~\eqref{eq:active} and~\eqref{eq:inactive}. 
Figure~\ref{fig:hidden_chirality}(b) shows the central reservoir effect: spin relaxation smooths the local imbalance between $\sigma^+$ and $\sigma^-$ reservoir populations and shifts their angular profiles relative to the pump. Because inactive excitons contribute only to the conservative blueshift whereas active excitons contribute both blueshift and gain, the real and imaginary parts of the effective condensate potential acquire different angular phases. This relative displacement breaks the equivalence between clockwise and counter-clockwise propagation, as shown in Fig.~\ref{fig:hidden_chirality}(c). In the absence of spin relaxation, or when the active and inactive reservoirs acquire identical angular phases, this non-Hermitian chirality vanishes.

To make this mechanism explicit, we adiabatically eliminate the reservoirs in the fast-reservoir limit~\cite{Chestnov2024,Pickup2021}. The condensate then obeys
\begin{equation}
i\hbar\partial_t\psi_\pm=\left[-\frac{\hbar^2\nabla^2}{2m_p}+V_\pm+U_\pm|\psi_\pm|^2+W_\pm|\psi_\mp|^2\right]\psi_\pm,
\label{eq:effective_gpe}
\end{equation}
where $V_\pm$, $U_\pm$, and $W_\pm$ are generally complex. The cross-spin nonlinear term $W_\pm|\psi_\mp|^2$ is generated by reservoir-mediated saturation and spin relaxation(see Suplementary Materials for details~\cite{[{See Supplementary Materials at [URL], which gives the details of the derivations}]SMBG}). The angular part of the complex potential can be written as
\begin{equation}
V_\pm=V_0(r)+V_2(r)\left[\tilde v_\pm\cos(2\theta+\tilde\chi_\pm)+i\bar v_\pm\cos(2\theta+\bar\chi_\pm)\right],
\label{eq:potential}
\end{equation}
where $\tilde v_\pm$ and $\bar v_\pm$ are real non-negative amplitudes. A finite phase offset
\begin{equation}
\delta\chi_\pm=\tilde\chi_\pm-\bar\chi_\pm\notin \{0,\pi\}
\end{equation}
produces a chiral non-Hermitian potential, since the maxima of the conservative and dissipative landscapes are no longer collinear.

At weak pump modulation the condensate can be expanded in angular modes of the radially symmetric trap. Annular polariton traps are often governed by a pair of fastest-growing modes with opposite angular momenta~\cite{Chestnov2024,Nalitov2019, cherotchenko2021optically}. Restricting to $m=\pm1$, we write
\begin{equation}
\psi_\pm(r,\theta,t)=h(r)\left[A^\pm_{+1}(t)e^{i\theta}+A^\pm_{-1}(t)e^{-i\theta}\right],
\end{equation}
where $h(r)$ is the radial mode profile. Projection of Eq.~\eqref{eq:effective_gpe} yields the frequency-normalized coupled-mode equation
\begin{equation}
i\dot{\bm A}=\hat L\bm A+\bm N,
\label{eq:cme}
\end{equation}
with $\bm A=(A^+_{+1},A^+_{-1},A^-_{+1},A^-_{-1})^\mathrm{T}$. The linear operator has block form
\begin{equation}
\hat L=\begin{pmatrix}
\hat L_+&0\\
0&\hat L_-
\end{pmatrix},\qquad
\hat L_\pm=\begin{pmatrix}
i\gamma_0&b_\pm\\
a_\pm&i\gamma_0
\end{pmatrix},
\label{eq:linear_matrix}
\end{equation}
where $\gamma_0$ is the net linear gain and $a_\pm$, $b_\pm$ are the scattering amplitudes between counter-rotating modes. All coefficients in Eq.~\eqref{eq:cme} are expressed in units of inverse time. The nonlinear terms read
\begin{align}
N^\pm_{\pm1}=&\;\eta_{\rm s}\left(|A^\pm_{\pm1}|^2+2|A^\pm_{\mp1}|^2\right)A^\pm_{\pm1}
\nonumber\\
&+\eta_{\rm c}\left[\left(|A^\mp_{\pm1}|^2+|A^\mp_{\mp1}|^2\right)A^\pm_{\pm1}
+A^\mp_{\pm1}\left(A^\mp_{\mp1}\right)^*A^\pm_{\mp1}\right],
\label{eq:nonlinear}
\end{align}
where $\eta_{\rm s}$ and $\eta_{\rm c}$ are complex self- and cross-spin nonlinear coefficients. Their imaginary parts describe gain saturation in the convention of Eq.~\eqref{eq:cme}, while their real parts describe nonlinear frequency shifts. 
The derivation of~\eqref{eq:cme} described in~\cite{SMBG} is based on non-Hermitian perturbation theory proposed for non-equilibrium exciton-polaritons condensates in~\cite{utesov2025universal}. The calculated parameters Eq.~\eqref{eq:cme} are listed in~\cite{parametersMode}.

\begin{figure}[t]
\centering
\includegraphics[width=1\columnwidth]{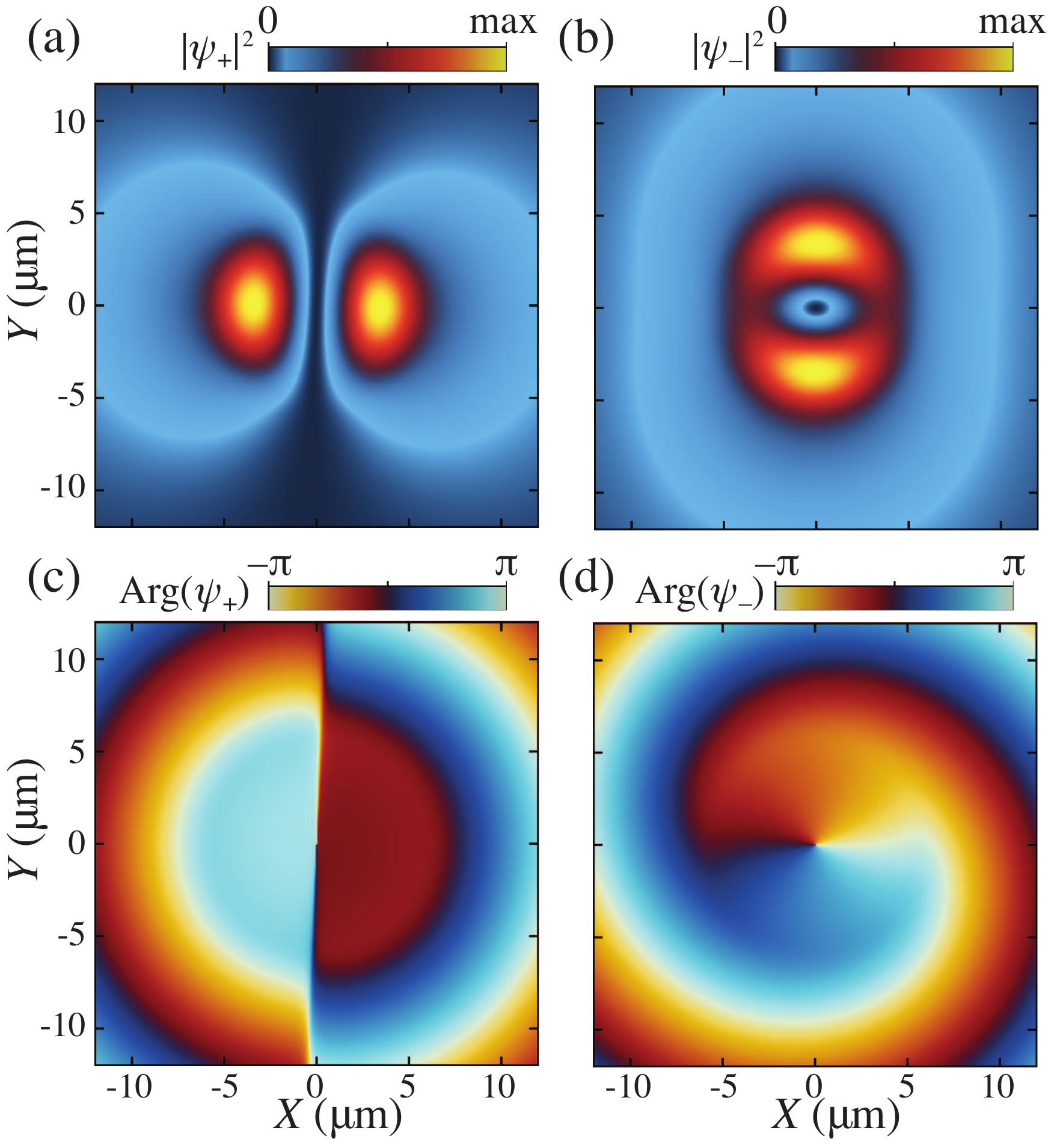}
\caption{Half-vortex state obtained from the full two-dimensional driven-dissipative Gross-Pitaevskii model near the EP of the $\sigma^-$ component. Panels (a) and (b) show the spin-resolved condensate densities $|\psi_+|^2$ and $|\psi_-|^2$, respectively; panels (c) and (d) show the corresponding phases. The $\sigma^+$ component forms a zero-winding dipole-like state, whereas the $\sigma^-$ component carries a $2\pi$ phase winding. The pump parameters are chosen close to the EP condition, with the notation of Eq.~\eqref{eq:rho}.}
\label{fig:half_vortex}
\end{figure}

The non-reciprocal nature of the chiral potential is encoded in $a_\pm\neq b_\pm$. The eigenvalues of each linear block are
\begin{equation}
\lambda^\pm_{1,2}=i\gamma_0\pm\sqrt{a_\pm b_\pm}.
\label{eq:eigenvalues}
\end{equation}
Thus an EP occurs when $a_\pm b_\pm=0$ while $\hat L_\pm\neq i\gamma_0\mathbb{I}$. At such a point the two eigenvalues and eigenvectors coalesce, and the surviving eigenvector is a purely chiral angular state. The EP condition can be reached by tuning $\rho$ and $\Delta\phi$: in Fig.~\ref{fig:hidden_chirality}(d), the upper map identifies the EP of the $\sigma^+$ block, while the lower map identifies the corresponding EP of the $\sigma^-$ block. Their displacement in parameter space shows that the hidden chirality can be addressed selectively in either circular polarization.

We confirm this picture using full two-dimensional simulations of Eqs.~\eqref{eq:gpe}-\eqref{eq:inactive} with parameters typical of GaAs-based microcavities~\cite{ParametersNote}. Close to the EP associated with the $\sigma^-$ polarization, the condensate forms a half-vortex state. Figure~\ref{fig:half_vortex}(a) shows that the $\sigma^+$ density has a dipole-like angular structure, while Fig.~\ref{fig:half_vortex}(c) confirms the absence of net phase winding in this component. By contrast, Fig.~\ref{fig:half_vortex}(b) shows the density of the $\sigma^-$ component and Fig.~\ref{fig:half_vortex}(d) displays its $2\pi$ phase winding. The comparison of Figs.~\ref{fig:half_vortex}(a)-\ref{fig:half_vortex}(d) demonstrates that hidden chirality can imprint vorticity selectively in one polarization.

The condensate angular momentum is quantified by the dimensionless expectation value
\begin{equation}
m_\pm=\frac{\int \psi_\pm^*\left(-i\partial_\theta\right)\psi_\pm\,r\,dr\,d\theta}{\int |\psi_\pm|^2\,r\,dr\,d\theta}.
\label{eq:angmom}
\end{equation}
For the two-mode expansion this reduces to
\begin{equation}
m_\pm=\frac{|A^\pm_{+1}|^2-|A^\pm_{-1}|^2}{|A^\pm_{+1}|^2+|A^\pm_{-1}|^2}.
\label{eq:angmom_modes}
\end{equation}

\begin{figure}[h]
\centering
\includegraphics[width=1\columnwidth]{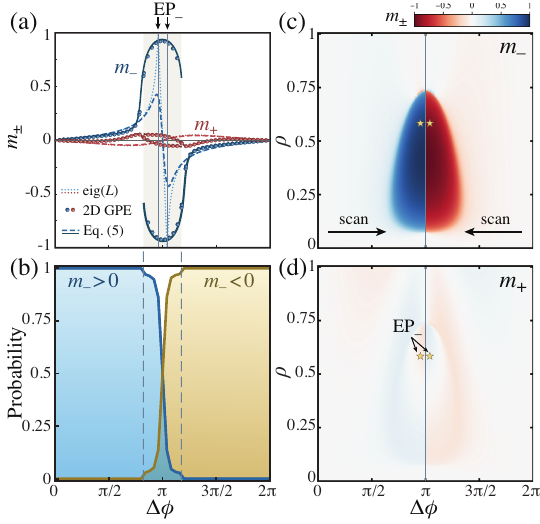}
\caption{Angular-momentum selection and nonlinear bistability near an exceptional point. (a) Spin-resolved mean angular momenta $m_+$ and $m_-$ as functions of the pump relative phase $\Delta\phi$ at fixed pump-modulation ratio $\rho$ corresponding to the EP of the $\sigma^-$ component. Linear eigenmode predictions, stable nonlinear coupled-mode solutions (just above -- dashed, and in the nonlinear regime -- solid lines), and full two-dimensional simulations are compared. Dashed gray regions indicate bistability. (b) Probability of selecting the positive- or negative-circulation branch from weak-noise initial conditions. (c,d) Phase diagrams of the steady angular momenta $m_-$~(c) and $m_+$~(d), where $m_\pm$ denotes the mean value fo angular momenta of the $\sigma_+$ and $\sigma_-$ polarized components in the $(\rho,\Delta\phi)$ plane. Near the $\sigma^-$ EP, the $\sigma^-$ component becomes strongly chiral, while the $\sigma^+$ component remains close to zero angular momentum, corresponding to half-vortex formation.}
\label{fig:angular_momentum}
\end{figure}

The dependence of this selection on the pump phase and amplitude is summarized in Fig.~\ref{fig:angular_momentum}. Figure~\ref{fig:angular_momentum}(a) compares the angular momenta predicted by the linear eigenmodes, the nonlinear coupled-mode model, and the full two-dimensional simulations as $\Delta\phi$ is scanned at fixed $\rho$ near the $\sigma^-$ EP. Near this EP, the fastest-growing linear eigenmode has $|m_-|\simeq1$, and nonlinear saturation preserves a strongly chiral $\sigma^-$ component while keeping $m_+$ close to zero. Figure~\ref{fig:angular_momentum}(b) shows that the nonlinear dynamics can be bistable, with weak-noise initial conditions selecting between the positive- and negative-circulation branches with probabilities that vary across the EP. Figure~\ref{fig:angular_momentum}(c) maps the resulting $m_-$ over the full $(\rho,\Delta\phi)$ plane, whereas Fig.~\ref{fig:angular_momentum}(d) shows the corresponding $m_+$ map. Together, Figs.~\ref{fig:angular_momentum}(c) and~\ref{fig:angular_momentum}(d) demonstrate that the half-vortex regime extends beyond the immediate vicinity of the EP.

\begin{figure}[h]
\centering
\includegraphics[width=1\columnwidth]{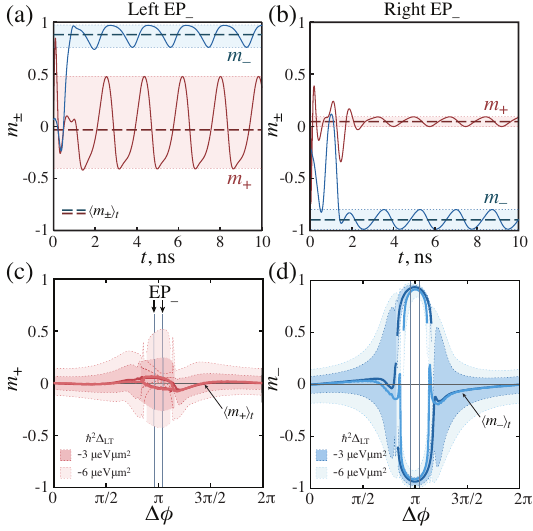}
\caption{Effect of TE-TM splitting on the hidden-chirality response. (a) Time evolution of the spin-resolved angular momenta on one side of the $\sigma^-$ EP in the presence of TE-TM coupling. (b) Corresponding time evolution on the opposite side of the same EP. The splitting induces fast spin-orbit oscillations between the two circular components. (c) Time-averaged $\langle m_+\rangle_t$ as a function of the pump relative phase $\Delta\phi$ for representative TE-TM strengths. (d) Time-averaged $\langle m_-\rangle_t$ over the same scan. The oscillation amplitude increases with splitting, but the time-averaged chiral response and half-vortex character remain robust for moderate TE-TM coupling.}
\label{fig:tmtm}
\end{figure}

We next examine whether the hidden-chirality response survives perturbations that lift the circular-polarization degeneracy. A relevant example is TE-TM splitting, which introduces the spin-orbit coupling~\cite{PhysRevB.106.235306, yulin2023spin}
\begin{equation}
i\hbar\partial_t\psi_\pm\big|_{\rm LT}=-\Delta_{\rm LT}\left(\hat p_x\mp i\hat p_y\right)^2\psi_\mp,
\label{eq:tmtm_realspace}
\end{equation}
with $\hat p_{x,y}=-i\hbar\partial_{x,y}$. In the $m=\pm1$ basis this term couples $A^+_{-1}$ and $A^-_{+1}$, giving
\begin{equation}
\hat L_{\rm LT}=\begin{pmatrix}
i\gamma_0&b_+&0&0\\
a_+&i\gamma_0&\Omega_{\rm LT}&0\\
0&\Omega_{\rm LT}&i\gamma_0&b_-\\
0&0&a_-&i\gamma_0
\end{pmatrix}.
\label{eq:tmtm_matrix}
\end{equation}
Here $\Omega_{\rm LT}$ is the projected TE-TM coupling, which may be complex in the dissipative basis. The TE-TM term modifies the full four-mode spectrum; nevertheless, for moderate splittings relevant to the parameters considered here, it primarily induces fast oscillations of $m_\pm(t)$ while leaving the time-averaged chiral response centered near the unperturbed EP positions. This is illustrated in Fig.~\ref{fig:tmtm}(a) for one side of the $\sigma^-$ EP and in Fig.~\ref{fig:tmtm}(b) for the opposite side, where the instantaneous angular momenta oscillate rapidly but preserve opposite mean signs. Figure~\ref{fig:tmtm}(c) shows that the time-averaged $\langle m_+\rangle_t$ remains close to the zero-winding component of the half-vortex, while Fig.~\ref{fig:tmtm}(d) shows that $\langle m_-\rangle_t$ retains a strong chiral response across the same parameter range. This explains why the half-vortex character remains visible in time-integrated observables even when instantaneous angular-momentum oscillations are present.

Experimentally, the required pump pattern can be generated by polarization-resolved spatial-light modulation or vector-beam shaping of a nonresonant annular excitation. The predicted hidden chirality can be tested by comparing the density and phase signatures in Figs.~\ref{fig:half_vortex}(a)-\ref{fig:half_vortex}(d) with angular-momentum scans of the type shown in Figs.~\ref{fig:angular_momentum}(a)-\ref{fig:angular_momentum}(d), for finite and suppressed reservoir spin relaxation, or equivalently by tuning the relative phase $\Delta\phi$ across the EP condition. The key signatures are the emergence of spin-selective winding, a strong asymmetry between counter-rotating angular harmonics, and the persistence of the time-averaged response under moderate TE-TM splitting. 
For the case of stronger TE-TM splitting (e.g. $\hbar^2\Delta_{\rm LT}\gtrsim 20\,\mu$eV$\mu$m$^{2}$ for our parameters), the two polarizations synchronize, their relative phase locks, and the oscillations cease. A detailed analysis of this synchronization regime is beyond the scope of this Letter and merits a separate investigation.

In conclusion, we have shown that a polariton condensate in a radially symmetric annular trap can acquire chirality through the combined action of spin relaxation and a two-component exciton reservoir. A spin-dependent but non-rotating pump modulation is converted into a relative displacement between the conservative and dissipative parts of the effective potential, producing non-reciprocal coupling between counter-rotating polariton modes. Near exceptional points this coupling becomes strongly chiral and enables the formation of half-vortex states. The robustness analysis indicates that the effect survives moderate TE-TM splitting and suggests reservoir engineering as a general route to programmable non-Hermitian chirality and spin-selective vorticity in driven condensates.

\begin{acknowledgments}
The work of I.Y.C. was supported by the Russian Science Foundation Grant No. 25-72-20029.
\end{acknowledgments}

\bibliographystyle{apsrev4-2}
\bibliography{bib_including_intro}

\end{document}


\author{A.N. Osipov}
\affiliation{School of Physics and Engineering, ITMO University, Saint Petersburg 197101, Russia}

\author{I.Y. Chestnov}
\affiliation{School of Physics and Engineering, ITMO University, Saint Petersburg 197101, Russia}

\author{P.G. Lagoudakis}
\affiliation{Hybrid Photonics laboratory, Skolkovo institute of Science and technology, Bolshoy Boulevard 30, Building 1, Moscow 121205,
Russia.}

\author{A.V. Yulin}
\affiliation{School of Physics and Engineering, ITMO University, Saint Petersburg 197101, Russia}

\title{ Supplementary materials for \\
\textit{Hidden chirality and half-vortex formation in exciton-polariton condensates}}

\maketitle

\tableofcontents

\section{Adiabatic elimination of the reservoirs}

We begin our theoretical derivations with the generalized driven-dissipative Gross–Pitaevskii model introduced in the main text, see Eqs.~(1):
\begin{subequations}
\label{2D GP with 2 res}
\begin{align}
    \label{eq.2DGPE}
    &i\hbar\partial_t \psi_{\pm} = \Big[-\frac{\hbar^2\bm{\hat{\nabla}}^2}{2 m_p}  + g(X_\pm^A + X_\pm^I) +\alpha|\psi_{\pm}|^2 + \frac{i\hbar}{2}(RX_{\pm}^A - \Gamma)\Big]\psi_\pm 
    , \\
    &\partial_t X_{\pm}^A = WX^I_{\pm} - (R|\psi_\pm|^2 + \Gamma_A)X_\pm^A - \Gamma_s (X_\pm^A - X_\mp^A), \\
    &\partial_t X_{\pm}^I = P_{\pm} - (W + \Gamma_I)X_\pm^I - \Gamma_s (X_\pm^I - X_\mp^I).
\end{align}
\end{subequations}
Assuming the relaxation dynamics of the reservoirs are fast, we can reduce the differential equations for the reservoirs to algebraic ones by setting $\partial_t X_\pm^I =0$ and $\partial_t X_\pm^A =0$. The resulting equations are linear with respect to the reservoir densities. 
To derive the equation for the polariton order parameter, we need to solve the algebraic equations for the reservoirs, $X^{A}$ and $X^{I}$, and express them as functions of $|\psi_{\pm}|^2$. Starting with the inactive reservoir, from Eq.~(\ref{2D GP with 2 res}c) we obtain:
\begin{equation}
    X_{\pm}^{I} = \hat T_1 \bm{P},
    \label{Dark reservoirs densities}
\end{equation}
where $\bm{X}^I=(X_{+}^{I}, X_{-}^{I})^\mathrm{T}$, $\bm{P}=(P_{+}, P_{-})^\mathrm{T}$ and 
\begin{equation}
    \hat T_1 = \begin{pmatrix} c^I_1 & c^I_2
    \\ c^I_2 &  c^I_1 
    \end{pmatrix},
\end{equation}
with $c^I_1 = \frac{W+\Gamma_I + \Gamma_s}{(W+\Gamma_I)(W+\Gamma_I + 2\Gamma_s)}$ and $c^I_2 = \frac{\Gamma_s}{(W+\Gamma_I)(W+\Gamma_I + 2\Gamma_s)}$. Similarly, from Eq.~(\ref{2D GP with 2 res}b), an expression for $X^A_{\pm}$ reads:
\begin{equation}
    \bm{X}^A = \hat{T}_2 \bm{X}^I,
    \label{X_A_general}
\end{equation}
where $\bm{X}^A=(X_{+}^{A}, X_{-}^{A})^\mathrm{T}$ and
\begin{equation}
    \hat{T}_2 = \frac{W}{\Delta}\begin{pmatrix} \Gamma_A + \Gamma_s + R |\psi_-|^2 & \Gamma_s
    \\ \Gamma_s &  \Gamma_A + \Gamma_s + R |\psi_+|^2 
    \end{pmatrix}.
\end{equation}
Here, the prefactor is given by the expression:
\begin{align}
    \frac{W}{\Delta} = \frac{W}{(\Gamma_A+\Gamma_s + R |\psi_-|^2)(\Gamma_A+\Gamma_s + R |\psi_+|^2) - \Gamma_s^2}.
\end{align}
Consequently, for the active reservoir, we obtain:
\begin{equation}
    \bm{X}^A = \hat{T}_2 \hat T_1 \bm{P}.
    \label{X_A_general_}
\end{equation}

By introducing the nonlinear complex potential
\begin{equation*}
    \bm{{\cal V}}=\begin{pmatrix} \mathcal{V}_{+} \\ \mathcal{V}_{-} \end{pmatrix}=\left( (g+\frac{i\hbar}{2}R) \hat{T}_2 +g \right)\hat T_1 \bm{P} + \begin{pmatrix} \alpha|\psi_{+}|^2 - \dfrac{i\hbar \Gamma }{2} \\[2pt] \alpha|\psi_{-}|^2 - \dfrac{i\hbar \Gamma }{2}
 \end{pmatrix}
\end{equation*}

one can write the equation for $\psi_{\pm}$ in the following form:
\begin{equation}
\label{2D GP no res}
i\hbar\partial_t \psi_{\pm} = \Big[-\frac{\hbar^2\bm{\hat{\nabla}}^2}{2 m_p}  +{\cal V}_{\pm} \Big]\psi_\pm .
\end{equation}
Note that this equation is nonlinear because the components of $\bm{{\cal V}}$ depend directly on the polariton densities, $|\psi_{\pm}|^2$.
While exact analytical expressions for $\bm{{\cal V}}$ can be derived, they are highly cumbersome. For our purposes, it is sufficient to analyze the dynamics under the assumption that the polariton densities remain low. This assumption holds well for pump powers close to the condensation threshold. Although this simplified model cannot describe the full dynamics when the pump significantly exceeds the threshold, it accurately captures the early stages of mode competition, where the nonlinearity can still be treated as a minor perturbation. 
Therefore, we expand the potential in Taylor series:
$$\bm{{\cal V}} \approx \bm{V} +\left( U_{+}|\psi_{+}|^2+W_{+}|\psi_{-}|^2, U_{-}|\psi_{-}|^2+W_{-}|\psi_{+}|^2  \right)^\mathrm{T},  $$
where $\bm{V}=(V_{+}, V_{-})^\mathrm{T}$ represents the effective linear potential, and $U_{\pm}=\frac{\partial {\cal V}_{\pm}}{\partial |\psi_{\pm}|^2}$ with $W_{\pm}=\frac{\partial {\cal V}_{\pm}}{\partial |\psi_{\mp}|^2}$ denote the nonlinear interaction coefficients, with the derivatives evaluated at zero density ($|\psi_{\pm}|^2=0$).
The equations for the order parameter then become:
\begin{equation}
i\hbar\partial_t\psi_\pm=\left[-\frac{\hbar^2{\bm\nabla}^2}{2m_p}+V_\pm+U_\pm|\psi_\pm|^2+W_\pm|\psi_\mp|^2\right]\psi_\pm.
\label{nonl_eq_taylor}
\end{equation}

Next, we demonstrate that the effective linear potential generated by the pump can exhibit chirality, even if the pump itself is non-chiral (in the sense discussed in the main text). The effective linear potential is given by:
\begin{align}
    & V_\pm = \left((gc_1^c + \frac{i\hbar R}{2}c_1^d)P_\pm + (gc_2^c + \frac{i\hbar R}{2}c_2^d)P_\mp \right)-i\hbar\Gamma/2,
    \label{V_eff}
\end{align}
where the expansion coefficients read:
\begin{eqnarray}
    \label{expansion coefficients}
    c_1^c  &=& Wc_1^{AI} + c_1^I, \quad c_2^c =  Wc_2^{AI} + c_2^I, \quad c_1^d = Wc_1^{AI}, \quad c_2^d = Wc_2^{AI}, \quad
    c_1^{AI} = (c_1^Ic_1^A + c_2^I c_2^A),  \\
    \quad c_2^{AI} &=& (c_2^Ic_1^A + c_1^I c_2^A), \quad c^A_1 = (\Gamma_A + \Gamma_s)/\left(\Gamma_A(\Gamma_A+2\Gamma_s)\right), \quad c^A_2 = \Gamma_s/\left(\Gamma_A(\Gamma_A+2\Gamma_s)\right). \nonumber
\end{eqnarray}

Importantly, the hidden chirality of the potential arises from two primary factors. First, the effective potential for each polarization has contributions from both circularly polarized pump components due to the presence of spin relaxation. Second, the complex phases of these two contributions differ. To satisfy the second condition, the coefficients in~\eqref{V_eff} must exhibit different ratios for self- and cross-polarization modulation, $c_1^c/c_1^d \neq c_2^c/c_2^d$. This requirement is naturally met for realistic, moderate spin-relaxation rates. Additionally, the dissipative and conservative contributions must also differ ($c_{1,2}^c \neq c_{1,2}^d$). This discrepancy is provided because the dark reservoir only contributes to the energetic blueshift, not the gain (as seen in~\eqref{expansion coefficients}, where $c_{1,2}^c = Wc_{1,2}^{AI}+c_{1,2}^I = c_{1,2}^d+c_{1,2}^I$ consists of terms stemming from the active reservoir expansion and the dark reservoir-induced blueshift, respectively). Consequently, a chiral effective potential can emerge even in optical traps formed by non-chiral pumping.

Next, we determine the effective nonlinear potentials from \eqref{nonl_eq_taylor}, which take the form:
\begin{subequations}
\begin{align}
    &U_\pm = \alpha -\left(g + \dfrac{i\hbar R}{2}\right)\left(c_1^{AI}P_\pm + c_2^{AI}P_\mp\right)RWc_1^A, \\
    &W_\pm =  \left(g + \dfrac{i\hbar R}{2}\right)\left(c_2^{AI}P_\pm + c_1^{AI}P_\mp\right)RWc_2^A.
\end{align}
\label{Nonlin_ampl}
\end{subequations}

\section{Linear coupled modes model}

To proceed with perturbation theory, we start with the linearized Eq.~\eqref{nonl_eq_taylor} and  split the  Hamiltonian into two parts:
\begin{equation}
    \hat{H} = -\frac{\hbar^2\hat{\nabla}^2}{2 m_p}  + V_{\pm} = \hat{H}_0 + \hat{H}_1,
\end{equation}
where the diagonal part $\hat{H}_0 = -\frac{\hbar^2\hat{\nabla}^2}{2 m_p} + V_0$
represents the effective linear Hamiltonian for exciton-polaritons confined in a symmetric annular trap, and 
$$
\hat{H}_1 = \begin{pmatrix}
    V_+-V_0 & - \Delta_{\mathrm{LT}}(\hat{p}_x - i\hat{p}_y)^2 \\
    - \Delta_{\mathrm{LT}}(\hat{p}_x + i\hat{p}_y)^2 & V_{-} -V_0
\end{pmatrix}
$$
corresponds to the part of the linear Hamiltonian arising from trap modulation and TE-TM splitting. Here {$V_0= \left((gc_1^c + \frac{i\hbar R}{2}c_1^d)p_0P(r) + (gc_2^c + \frac{i\hbar R}{2}c_2^d)p_0P(r) \right) - i\hbar\Gamma/2$ is angular symmetric part of the potential}. In the subsequent analysis, we treat the interaction Hamiltonian $\hat{H}_1$ perturbatively, assuming both shallow angular modulation of the pump and weak TE-TM splitting.

The polariton fields $\psi_{\pm}$ can be expanded in terms of the eigenmodes of the operator $\hat{H}_0$ \cite{cherotchenko2021optically}:
\begin{equation}
    \psi_{\pm} (\bm{r}, t) = \sum_{m, l} A_{m, l}^{\pm} (t)h_{m,l}(r)e^{im\theta} ,
    \label{Mode expansion suppl_}
\end{equation}
where $A_{m,l}^{\pm}(t)$ are the time-dependent expansion coefficients (hereafter referred to as mode amplitudes), $m$ denotes the azimuthal quantum number (angular index), and $l$ enumerates the radial nodes. The radial wavefunctions $h_{m,l}(r)$ correspond to the solutions of the eigenvalue equation:
\begin{equation}
    \left(\frac{-\hbar^2}{2 m_p}\left(\frac{1}{r}\partial_r(r\partial_r\cdot) - \frac{m^2}{r^2}\right) + P(r)p_0\frac{(g(W+\Gamma_A)+i\hbar WR/2)}{\Gamma_A(W+\Gamma_I)} - i\hbar\Gamma/2\right)h_{m,l}(r) = (E_{m,l} -i\hbar\gamma_{m,l}) h_{m,l}(r). 
\end{equation}

We focus on a parameter regime where the fundamental radial modes ($l=0$) with angular momentum $m=\pm 1$ experience the highest gain from the incoherent exciton reservoir. Consequently, in the leading-order approximation, the system dynamics are strictly governed by the evolution of these four lowest-energy modes. Any corrections stemming from the non-resonant excitation of higher-order modes can be treated perturbatively. As detailed in the main text, we truncate the polariton field expansion to these four principal modes ($m = \pm 1$): 
\begin{equation}
    \psi_{\pm} (\bm{r}, t) = A_{+1}^{\pm} (t)h(r)e^{i\theta} + A_{-1}^{\pm} (t)h(r)e^{-i\theta}.
    \label{Mode expansion suppl}
\end{equation}
Here, we utilize the symmetry $h_{m,l}(r)=h_{-m,l}(r)$. For the sake of brevity, we drop the radial index $l=0$ and denote $h_{\pm 1, 0}(r)$ simply as $h(r)$ throughout the remainder of this text. Notably, these selected eigenmodes are spatially localized.

To derive the ordinary differential equations for the amplitudes $A$, we apply a standard perturbation theory technique based on the condition that the perturbation does not result in the infinite growth of the small correction to the solution (\ref{Mode expansion suppl}). We substitute the ansatz~\eqref{Mode expansion suppl}, treating $A^{\pm}_{\pm 1}$ as slowly varying functions of time, and require that the resulting equations be orthogonal to the eigenfunctions of the adjoint operator $\hat{H}_0^{\dagger}$, which correspond to the complex conjugated eigenvalues $E_{m,l} + i\hbar\gamma_{m,l}$.
Since the linear non-Hermitian operator is symmetric ($\hat{H}_0 = \hat{H}_0^\mathrm{T}$), the following relationship holds for any eigenfunction $\psi$~\cite{utesov2025universal}:
\begin{equation}
    \hat{H}_0 \psi = \lambda\psi \rightarrow  \hat{H}_0^\dagger \psi^* = \hat{H}_0^* \psi^* = \lambda^*\psi^*.
\end{equation}
Therefore, $\psi^*$ is exactly the eigenfunction of the adjoint operator. Projecting onto the eigenfunctions of $\hat{H}_0^{\dagger}$ is mathematically equivalent to taking the scalar product with the complex conjugate of the eigenfunctions of $\hat{H}_0$.


The Hamiltonian describing the effect of TE-TM splitting and the influence of the axially non-symmetric effective potential is given by:
\begin{equation}
    \hat{H}_1\psi_\pm = \left((gc_1^c + \frac{i\hbar R}{2}c_1^d)( p_2^\pm e^{i2\theta} + \mathrm{c.c.}) + (gc_2^c + \frac{i\hbar R}{2}c_2^d)(p_2^\mp e^{i2\theta} + \mathrm{c.c.}) \right)\psi_\pm  - \Delta_{\mathrm{LT}}(\hat{p}_x\mp i\hat{p}_y)^2\psi_{\mp}.
    \label{H_1}
\end{equation}
By applying the orthogonality condition, we extract the equations for the amplitudes:
\begin{subequations}   
\begin{align}
    &i\partial_t A_{+1}^+ = i\gamma_0 A_{+1}^+ +  (\xi_1p_2^+ + \xi_2p_2^-)  A_{-1}^+ ,  
    \label{eq +1+ initial}\\ 
    &i\partial_t A_{-1}^+ =  i\gamma_0 A_{-1}^+  +  (\xi_1(p_2^+)^* + \xi_2(p_2^-)^*)  A_{-1}^+ + \Omega_{\mathrm{LT}}A_{+1}^-,\\ 
    &i\partial_t A_{+1}^- =  i\gamma_0 A_{+1}^- +  (\xi_1p_2^- + \xi_2p_2^+)  A_{-1}^- + \Omega_{\mathrm{LT}}A_{-1}^+,\\
    &i\partial_t A_{-1}^- =  i\gamma_0 A_{-1}^-+  (\xi_1(p_2^-)^* + \xi_2(p_2^+)^*)  A_{+1}^-. 
    \label{eq -1- initial}
\end{align}
\end{subequations}

Thus, modes with the same polarization couple to each other with an interaction strength proportional to the amplitude of the second angular harmonic of the potential. This coupling mechanism appears directly from azimuthal orthogonality: the $e^{\pm 2i\theta}$ terms in~\eqref{H_1} yield non-vanishing matrix elements $H_{m_1m_2}$ only at $|m_1 - m_2| = 2$. Similarly, the TE-TM splitting couples the $\sigma_+$-polarized mode ($m=+1$) with the $\sigma_-$-polarized mode ($m=-1$), which is consistent with previously reported findings \cite{PhysRevB.106.235306, yulin2023spin}.

For convenience, we divide the equations by $\hbar$ and shift the energy frame via the transformation $A_{\pm 1}^\pm = e^{-i{E_0}t/\hbar}$, where $E_0$ is the unperturbed energy of the states with unit angular momentum $m =\pm 1$. The coefficients $\xi_1 = P_0(\frac{g}{\hbar}c_1^c + \frac{iR}{2}c_1^d)I_p$ and $\xi_2 = P_0(\frac{g}{\hbar}c_2^c + \frac{iR}{2}c_2^d)I_p$ denote the contributions from the non-Hermitian potential generated by the pump of the same  and the opposite polarization, respectively. Here, $I_p = \int h(r)^2P(r)rdr/I_0$ represents the spatial overlap integral between the vortex modes ($m= \pm 1$) and the radial pump profile, $I_0 = \int h(r)^2rdr$ is the normalization constant, and $\Omega_{\mathrm{LT}} = -\hbar\Delta_{\mathrm{LT}}\int h(r)(\partial_r)(\partial_r + \frac{1}{r})h(r)rdr/I_0$ defines the effective TE-TM splitting between the vortex modes. The mode dynamics equations can then be rewritten in the format utilized in the main text:
\begin{subequations}
\begin{align}
    &i\partial_t A_{+1}^{+} = i\gamma_0 A_{+1}^{+}  + b_+ A_{-1}^{+} , \label{linear modes int eq 1}\\ 
    &i\partial_t A_{-1}^{+} = i\gamma_0 A_{-1}^{+} + a_+ A_{+1}^{+} + \Omega_{\mathrm{LT}}A_{+1}^{-},   \\ 
    &i\partial_t A_{+1}^{-} = i\gamma_0 A_{+1}^{-} + b_- A_{-1}^{-} + \Omega_{\mathrm{LT}}A_{-1}^{+}, \\ 
    &i\partial_t A_{-1}^{-}= i\gamma_0 A_{-1}^{-} + a_- A_{+1}^{-} , \label{linear modes int eq 4}
\end{align}
\label{linear mode equations}
\end{subequations}
where the effective couplings are defined:
\begin{subequations}
    \label{couplings}
    \begin{align}
        &a_\pm = \xi_1(p_2^\pm)^* + \xi_2(p_2^\mp)^*, \\
        &b_\pm = \xi_1p_2^\pm + \xi_2p_2^\mp .
    \end{align}
\end{subequations}

The exceptional points can be easily identified from~\eqref{linear mode equations}. Notice that the linear operator on the right-hand side has a Jordan block form if any of the coupling coefficients, $a_\pm$ or $b_\pm$, vanish. The exceptional points associated with different $\pm$ indices correspond to broken chirality in respective polarization. Meanwhile, the difference between exceptional points where either $a=0$ or $b=0$ dictates the direction of chirality breaking, ultimately favoring either an $m=+1$ or $m=-1$ vorticity direction. Let us illustrate this physical consequence with a concrete example where $b_+ = 0$, implying $p_2^- = -(\xi_1/\xi_2)p_2^+$. This condition yields the following expressions for the remaining couplings:
\begin{subequations}
\label{other couplings}
\begin{align}
    & b_- = -p_2^+\left(\frac{\xi_1^2-\xi_2^2}{\xi_2}\right), \\
    & a_+ = (p_2^+)^*\left(\frac{\xi_1\xi_2^*-\xi_1^*\xi_2}{\xi_2^*}\right) = (p_2^+)^*\left(\frac{2i\operatorname{Im}(\xi_1\xi_2^*)}{\xi_2^*}\right),\\
    & a_- = -(p_2^+)^*\left(\frac{|\xi_1|^2-|\xi_2|^2}{\xi_2^*}\right).
\end{align}
\end{subequations}
Equations~\eqref{other couplings} demonstrate that all couplings except $b_+$ remain non-zero, provided that $\xi_1$ and $\xi_2$ possess different complex phases and absolute values. If we rewrite the linear mode interaction equations~\eqref{linear modes int eq 1} in a matrix form, $\partial_t\vec{A}=\hat{H}_L\vec{A}$, it becomes transparent that the Hamiltonian $\hat{H}_L$ behaves as a Jordan cell within one polarization. Specifically, the eigenvector $\vec{e}_1 = [0,1,0,-\Omega_{\mathrm{LT}}/b_-]^\mathrm{T}$ and generalized eigenvector $\vec{e}_2 = [1,0,-\Omega_{\mathrm{LT}}/(a_-b_-+\Omega_{\mathrm{LT}}^2),0]^\mathrm{T}$ satisfies $(\hat{H}_L-i\gamma_0)\vec{e}_1 = 0$ and $(\hat{H}_L-i\gamma_0)\vec{e}_2 = a_+\cdot\left(b_-a_-/((b_-a_-+\Omega_{\mathrm{LT}}^2)\right)\vec{e}_1$ forming generalized eigen subspace. For small LT splittings, these modes predominantly represent the $m =-1$ and $m=+1$ states in the $\sigma_+$ polarization. Consequently, in the mode interaction model~\eqref{linear mode equations} where $b_+ = 0$ and $a_+\neq 0$, the scattering between $m =\pm 1$ modes in the $\sigma_+$ polarization becomes strictly unidirectional (from $m=+1$ to $m=-1$). This asymmetric scattering forces the polariton density to transfer entirely from the $m=+1$ mode into the $m=-1$ mode, leading to effective linear growth with time of the $m=-1$ mode and concurrent decay of the $m=+1$ mode (which mathematically corresponds to the solution of an ODE featuring a Jordan matrix). Therefore, the system preferentially forms an $m=-1$ vortex in the $\sigma_+$ polarization, which is a manifestation of broken chirality. In contrast, in the $\sigma_-$ polarization, both couplings remain non-zero ($a_-\neq 0$, $b_-\neq 0$). This means chirality is not completely broken in the $\sigma_-$ polarization, allowing for comparable occupancies of the $m=\pm 1$ states. Thus, the four distinct exceptional points map to two polarizations and two effective chirality-breaking directions.

Let us note that the angular modulation of the pumping with the first harmonic $p_1^\pm e^{\pm i\theta}$ could also create the required chiral interaction. However, the mechanism proposed here relies on the coupling between the two vortex modes with angular momenta $m=\pm 1$. The second harmonic $p_2^\pm e^{\pm 2i\theta}$ thus provides a direct coupling at the first order in the modulation amplitude. In contrast, the first harmonic $p_1^\pm e^{\pm i\theta}$ couples the $m=\pm1$ modes only indirectly via non-resonant excitation of the intermediate $m=0$ mode. Adiabatic elimination of the $m=0$ mode yields an effective coupling between the $m=\pm1$ states which is of the second order of perturbation theory and scales as $(p_1^\pm)^2/\Delta\omega_{0,1}$, where $\Delta\omega_{0,1}$ is the eigenfrequency detuning between the $m=0$ and $m=\pm1$ modes. Thus, these indirect couplings are typically much weaker than the direct coupling induced by the second harmonics.

Let us also remark that in the presence of two angular harmonics in the trapping potential the formation of polariton half-vortices is possible without spin relaxation. The reason for this is that two angular harmonics with appropriate mutual phase break the axial symmetry of the system explicitly. This effect is out of the scope of the present paper and will be considered somewhere else.

\section{Nonlinear part of the coupled mode equations}

To leading order, the nonlinear effects are governed by the cubic terms in Eq.~\eqref{nonl_eq_taylor}. We employ the same perturbative approach developed in the previous section to derive the generalized equations for the mode amplitudes in the presence of this nonlinearity. By substituting the mode expansion~\eqref{Mode expansion suppl} into the general expression for the cubic nonlinear terms, $|\psi_{\sigma_1}|^2\psi_{\sigma_2}$, we obtain:
\begin{equation}
    |\psi_{\sigma_1}|^2\psi_{\sigma_2} = (A^{\sigma_1}_{+1}e^{i\theta} + A^{\sigma_1}_{-1}e^{-i\theta})(A^{\sigma_1}_{+1}e^{i\theta} + A^{\sigma_1}_{-1}e^{-i\theta})^*(A^{\sigma_2}_{+1}e^{i\theta} + A^{\sigma_2}_{-1}e^{-i\theta})|h(r)|^2h(r).
\end{equation}

By collecting the corresponding nonlinear coefficients and projecting them using the orthogonality condition discussed earlier, we evaluate the spatial overlap integrals providing mode equations nonlinearities in the form:
\begin{eqnarray}
   N^{\pm}_m &=&  (\eta'+i\eta'')(|A^{\pm}_m|^2 + 2|A^{\pm}_{-m}|^2)A^{\pm}_m + (\tilde\eta'+i\tilde\eta'')\left((|A^{\mp}_m|^2 + |A^{\mp}_{-m}|^2)A^{\pm}_m + A^{\mp}_m (A^{\mp}_{-m})^* A^{\pm}_{-m}\right),
\end{eqnarray}

where
\begin{subequations}
\begin{align}
    &\eta' = \alpha I_{nl}^0 -gP_0\left(c_1^{AI} + c_2^{AI}\right)RWc_1^AI_{nl}^P,\\
    & \eta'' =  -\frac{i\hbar R}{2}P_0\left(c_1^{AI} + c_2^{AI}\right)RWc_1^AI_{nl}^P, \\
    &\tilde\eta' =  -gP_0\left(c_1^{AI} + c_2^{AI}\right)RWc_2^AI_{nl}^P,\\
    & \tilde\eta'' =  -\frac{i\hbar R}{2}P_0\left(c_1^{AI} + c_2^{AI}\right)RWc_2^AI_{nl}^P. \\  
\end{align}
\end{subequations}
Here, nonlinear spatial overlap integrals are defined as $I_{nl}^P = \int P(r)|h(r)|^2 h(r)^2r dr/I_0$ and $I_{nl}^0 = \int |h(r)|^2 h(r)^2r dr/I_0$. The nonlinear terms with $\eta'$, $\eta''$ and $\tilde\eta'$, $\tilde\eta''$ directly come from the mode expansion expansion and integration of the nonlinear terms $U_\pm|\psi_\pm|^2\psi_\pm$ and $W_\pm|\psi_\mp|^2\psi_\pm$ of Eq.~\eqref{nonl_eq_taylor} with $U_\pm,\,W_\pm$ defined in \eqref{Nonlin_ampl}.
By grouping the coefficients at the terms $(|A^{\pm}_m|^2 + 2|A^{\pm}_{-m}|^2)A^{\pm}_m$ and $\left((|A^{\mp}_m|^2 + |A^{\mp}_{-m}|^2)A^{\pm}_m + A^{\mp}_m (A^{\mp}_{-m})^* A^{\pm}_{-m}\right)$, we recover the nonlinear expressions for the mode equations in the form presented in Eq.~(11) of the main text.

\section{The mode equations steady states}

To provide better understanding of the numerical simulations presented in the main text, it is instructive to find all stationary solutions of Eq.~(9) from the main text. To achieve this, we identify the corresponding fixed points by substituting the ansatz $A^{+}_{\pm 1}=\alpha^{+}_{\pm 1} e^{i\omega^{+}t}$, $A^{-}_{\pm 1}=\alpha^{-}_{\pm 1} e^{i\omega^{-}t}$ into Eq.~(9) of the main text, setting $\partial_t \vec{\alpha} = 0$, and subsequently solving the resulting system of nonlinear algebraic equations using homotopy-based numerical packages.

We study the fixed point solutions as a function of the phase difference $\Delta \phi$. We characterize each state by its angular momentum, evaluated in either the $\sigma_{+}$ or $\sigma_{-}$ polarization basis treating it as primary physical observables.

\begin{figure}[!tb]
\centering
\includegraphics[width=0.75\linewidth]{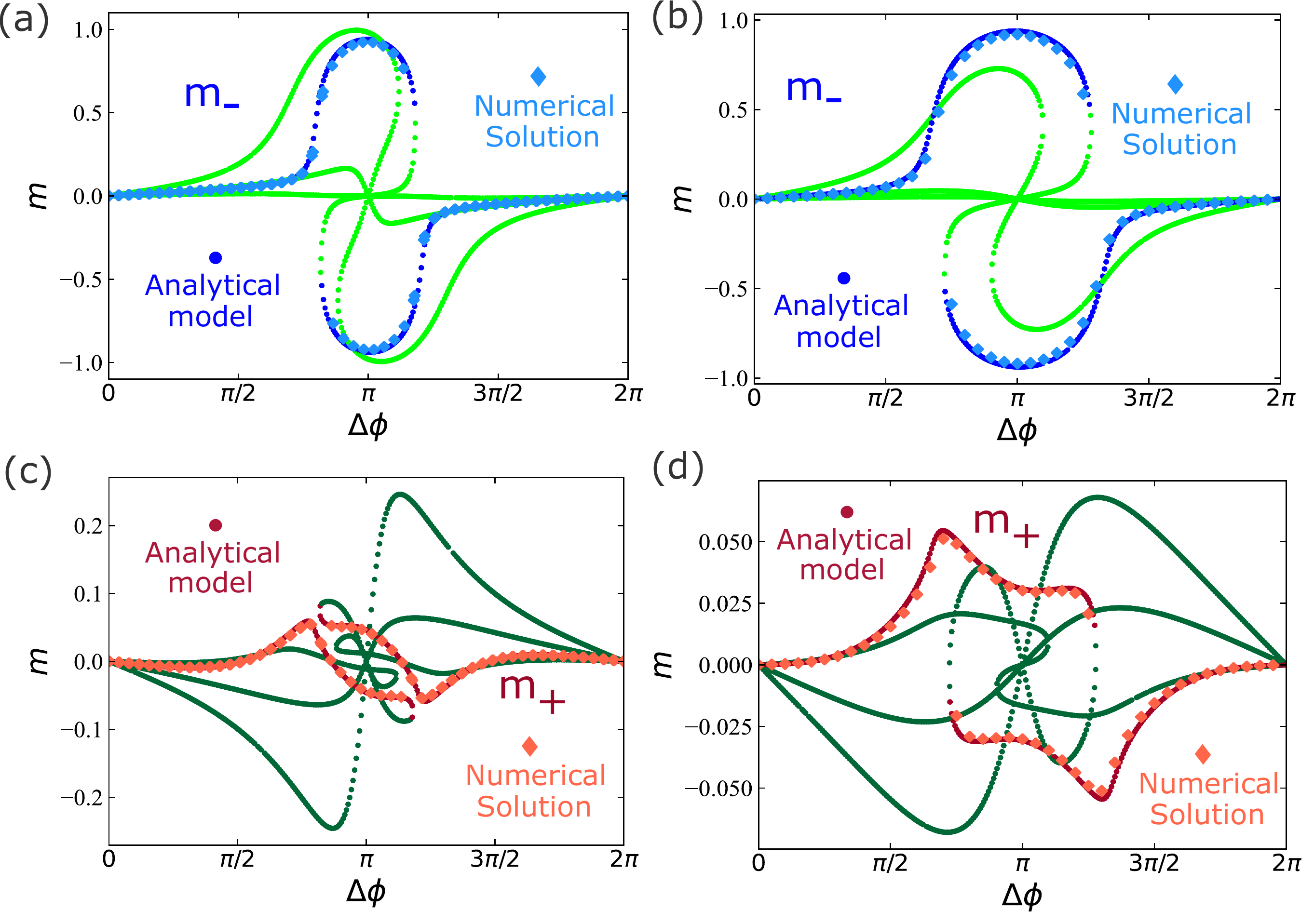}
\caption{Dependence of the angular momenta of numerically found steady states on the polarized pumps' modulation phase shifts $\Delta \phi$, for $\rho = \left| p_2^-/p_2^+\right| = \rho_{EP}$ in (a) and (c) and for $\rho = 0.25$ in (b) and (d). Panels (a) and (b) show the angular momentum $m_-$ of the $\sigma_-$ polarization, while panels (c) and (d) show the angular momentum $m_+$ of the $\sigma_+$ polarization. Dotted curves represent the numerically found steady states of the effective analytical mode interaction model~(9) from the main text. Light and dark green dots correspond to unstable fixed points, whereas red and blue dots correspond to stable states. The overlaid diamonds indicate the angular momenta extracted directly from the numerical integration of the Gross–Pitaevskii equation (Eq.~(1) from the main text). }
\label{FigS1}
\end{figure}

The resulting bifurcation diagrams are presented in Fig.~\ref{FigS1}, which displays the angular momenta of the calculated steady states for the effective mode equations (Eq.~(9) from the main text). Panels (a) and (c) correspond to a pump modulation depth of $\rho = \left| p_2^-/p_2^+\right| = \rho_{EP}$, illustrating the bifurcation diagrams in the $(m_{-},\Delta \phi)$ and $(m_{+},\Delta \phi)$ parameter spaces, respectively. Panels (b) and (d) present the identical analysis for $\rho=0.25$ which is away from exceptional point position. The linear stability of these solutions is verified by deriving linear equations for small-amplitude perturbations superimposed on the background of the stationary states and subsequently analyzing the eigenvalues governing their temporal evolution.

As observed in both scenarios ($\rho=\rho_{EP}$ and $\rho=0.25$), there are exactly two stable solution branches, represented by blue curves in panels (a,b) and red curves in panels (c,d). Additionally, there are six unstable branches, designated by green curves across all panels. Notably, both stable branches disappear as $\Delta\phi$ varies by colliding with a common unstable branch, annihilating via saddle-node bifurcations. Consequently, the two distinct stable branches are continuously connected by an intermediate unstable branch (see Fig.~\ref{FigS1}).
Because the parameter domains of these stable branches overlap, the system exhibits clear regions of bistability (see Fig.~\ref{FigS1}). Within these regions, the stable states belonging to different branches possess angular momenta of opposite signs.

Remarkably, the mode interaction effective model (Eq.~(9) in the main text) accurately reproduces the dynamics of the full 2D Gross-Pitaevskii simulations. For direct comparison, the steady-state angular momenta extracted from the full spatial simulations are superimposed on Fig.~\ref{FigS1} as solid diamonds. One can clearly see that the full simulations data perfectly follow the solutions of the simplified mode model.

\section{The effect of cross-polarization nonlinear interaction}

\begin{figure}[!tb]
\centering
\includegraphics[width=0.75\linewidth]{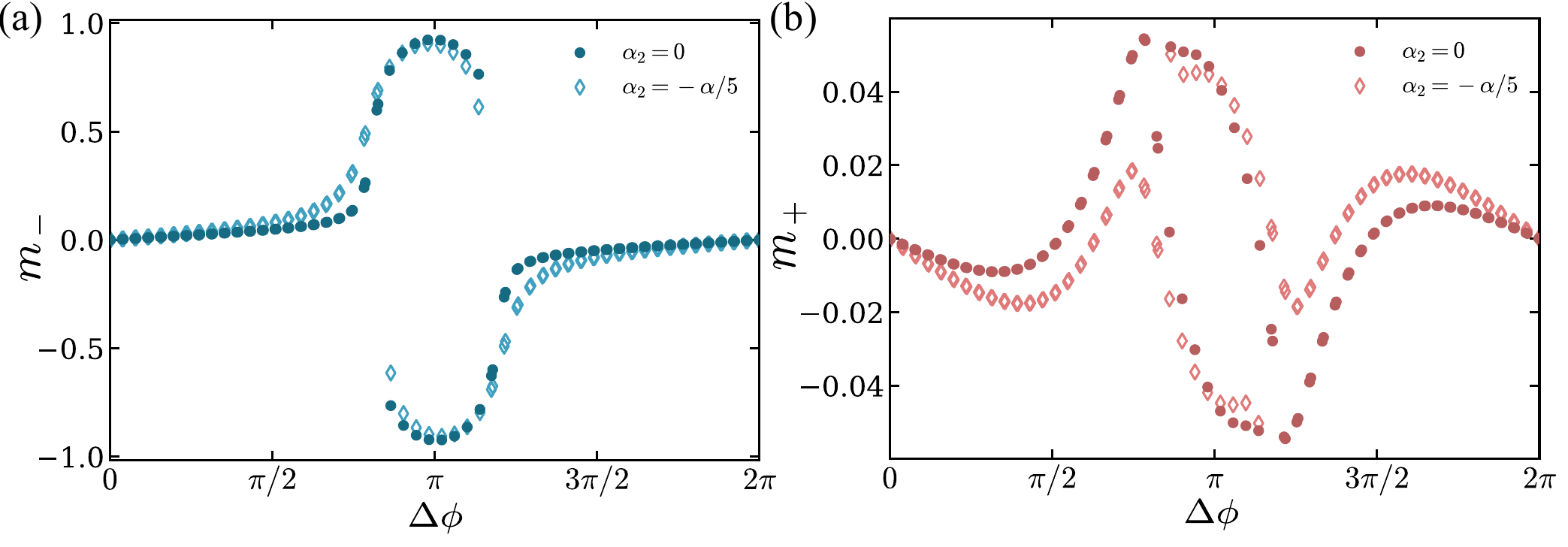}
\caption{Comparison of the stationary state circular polarization degrees $m_-$ (a) and $m_+$ (b) as a function of the phase difference $\Delta\phi$, calculated with and without the cross-polarization modulation term. Solid circles represent the simplified model ($\alpha_2 = 0$), while hollow diamonds represent the generalized model incorporating an experimentally realistic cross-polarization modulation ($\alpha_2 = -\alpha/5$).}
\label{FigS2}
\end{figure}

In the main text, we considered a minimal realistic model configuration, required to observe predicted effects. In general, an additional cross-polarization nonlinear phase modulation term may be present in GP equation (1a) or polariton condensates from the main text:
\begin{equation}
    i\hbar\partial_t \psi_{\pm} = \left[-\frac{\hbar^2\bm{\hat{\nabla}}^2}{2 m_p} + g(X_\pm^A + X_\pm^I) + \alpha|\psi_{\pm}|^2 + \frac{i\hbar}{2}(RX_{\pm}^A - \Gamma) + \alpha_2|\psi_\mp|^2 \right]\psi_\pm,
\end{equation}
where $\alpha_2|\psi_\mp|^2$ describes cross-polarization nonlinear phase modulation. However, this term is typically weak ($\alpha_2 \approx -\alpha/5$~\cite{PhysRevB.103.155302}) and yields no new physical effects, as reservoir-mediated cross-polarization modulation is already present in our model due to the spin relaxation. Thus, this interaction does not play a qualitative role and only slightly quantitatively shifts the predicted observables. To verify this, we performed numerical simulations of Eq.~(1) from the main text with the inclusion of $\alpha_2$ terms, using the same parameters as in Fig.~3. As shown in Fig.~\ref{FigS2}, the $\alpha_2$ term does not noticeably alter the results, fully justifying its omission for simplicity.

\bibliography{bib_including_intro}